\documentstyle[floats,aps,prb,epsf,twocolumn]{revtex}

\begin{document}
\draft


\twocolumn[\hsize\textwidth\columnwidth\hsize\csname
@twocolumnfalse\endcsname


\title{
On-site magnetization in open antiferromagnetic chains: a classical analysis
versus NMR experiments in a spin-1 compound}


\author{S. Botti$^1$, A. Rosso$^2$, R. Santachiara$^1$ and F. Tedoldi$^{1,}$\footnote{e-mail: tedoldi@unipv.it}}


\address{$^1$INFM - Department of Physics "A. Volta", Via Bassi 6, I-27100
Pavia, Italy}


\address{$^2$CNRS-Laboratoire de Physique Statistique\\
Ecole Normale Sup{\'{e}}rieure,
24, rue Lhomond, 75231 Paris Cedex 05, France}


\date{\today}
\maketitle
\widetext
\begin{abstract}


The response of an open spin chain with isotropic antiferromagnetic
interactions to a {\it uniform} magnetic field is studied by classical 
Monte Carlo simulations. It is observed how the induced on-site 
magnetization is {\it non uniform}, due to the occurrence of edge 
staggered terms which decay exponentially over a distance equal to
the zero field correlation length of the infinite chain. The total 
magnetic moment associated to each staggered term is found to be about 
half of the original single-spin magnitude and to decrease as the inverse
of temperature (i.e. to behave as a Curie-like moment).
The numerical results are compared to recent NMR findings in spinless-doped
Y$_{2}$BaNiO$_{5}$; the remarkable agreement found shows that, for 
temperatures  above the Haldane gap, the classical approach gives a correct
picture of the boundary effects observed in the Heisenberg $S$=1 chain.









\end{abstract}
\pacs {PACS numbers: 75.10.Jm, 75.40.Mg, 75.50.Ee, 76.60.-k}
] \newpage 


\narrowtext



Because of the richness of their phase diagrams, Heisenberg spin systems
with low-dimensional antiferromagnetic interactions (HAF's) are presently 
attracting strong interest. Magnetic correlations in spatially-homogeneus
HAF's have been probed dynamically, in Fourier-space, by neutron scattering,
but can hardly be visualized experimentally in a static real-space picture.
Translational invariance, infact, makes each site equivalent to the others, 
thus preventing any spatial oscillation of the spin direction. In recent 
experiments, the correlation properties of low-dimensional HAF's have been 
investigated through NMR imaging of the spin polarization induced by a
uniform field around non-magnetic defects that break the translational 
invariance. Relevant results have been obtained in copper oxide two-dimensional 
compounds\cite{Bobroff,Julien}, in spin ladders\cite{Fujiwara}, in half-integer%
\cite{Takigawa} and integer spin chains\cite{Tedoldi}.




In $S=1$ chains in particular, the local magnetization $\langle
S_{i}^{z}\rangle $ has been
resolved {\it site by site}, through $^{89}$Y NMR in Mg-doped
Y$_{2}$BaNiO$_{5}$
\cite{Tedoldi}, for temperatures ranging from $T\simeq 0.35 J/k_B$ to $T\simeq
J/k_B$, being $J$ the
AF exchange constant. $\langle S_{i}^{z}\rangle$ shows an alternate
component which is maximum
around impurities (i.e. close to chain boundaries) and vanishes
exponentially over a distance
equal to the zero-field correlation length of the bulk. Boundary staggered
defects with total
spin $S=1/2$ and a size of the order of the bulk correlation length are
actually expected,
for $S=1$ HAF chains, in the limit $T\rightarrow 0$. At very low
temperatures in fact,
the magnetic properties of these systems are controlled by edge-induced
triplet states%
\cite{Kennedy1,Kennedy2}, in which $\langle S_{i}^{z}\rangle$ shows the
profile described
above\cite{Miya,White,Sore1,Polizzi}. This $T\simeq 0$ argument however, can
hardly be used
by itself in order to explain the experimental evidences in Ref. 5, since -
on increasing
temperature - the excited states above the Haldane gap $\Delta_H\simeq0.4J$
should be taken
into account (in open $S=1$ chains $\Delta_H$ corresponds to the separation
between the second
and the third lowest lying energy levels\cite{Kennedy1}). A finite
temperature analysis, in
which all the excited states are correctly treated, was recently carried out
by Alet and
S\o rensen\cite{Alect}, using Quantum Monte Carlo techniques. Excellent
agreement with the
experimental data derived from NMR imaging\cite{Tedoldi} has been obtained.



\begin{figure}[b!]
\vspace{-10mm}
\begin{center}
\epsfxsize=80mm
 $$\epsffile{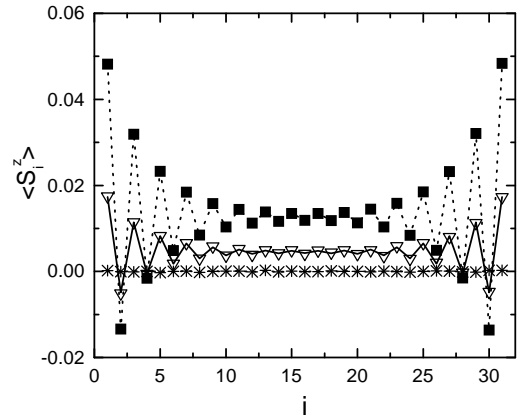}$$
\vspace{-6cm}
\caption{Magnetization profiles at $T=150$~K, obtained by classical Monte Carlo in 
a $S=1$ chain of 31 sites with $J/k_B$=285~K. Different symbols correspond to different 
applied fields: $H=0$ (stars), $H=5$ Tesla (triangles) and $H=14.1$ Tesla (squares).}
\label{fig1}
\end{center}
\end{figure}


\begin{figure*}[t!]
\vspace{-1cm}
\centerline{\epsfxsize=140mm \epsfbox{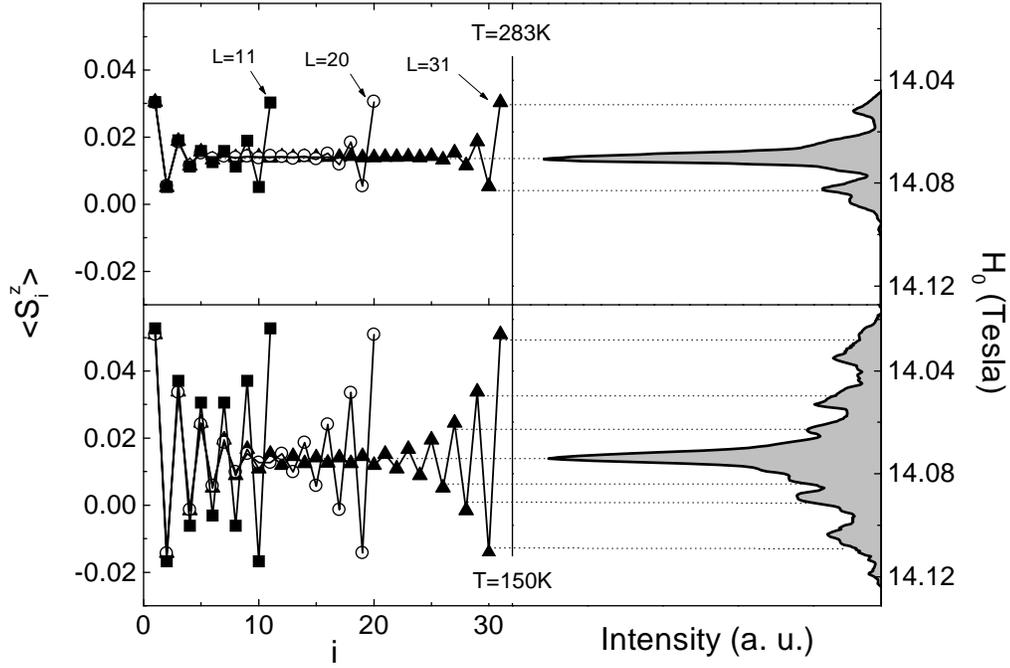}}
\vspace{-10cm}
\caption{Left side: magnetization profiles $\langle S_{i}^{z}\rangle$ in a
field  $H = 14.1$ Tesla for $S=1$ chains with $J/k_B=285K$ and various lengths 
(from classical Monte Carlo). Solid lines are fittings of the data according to 
the phenomenological model described by Eq. (\ref{model.eq}). On the right side 
$^{89}$Y NMR spectra in doped HAF chain Y$_2$BaNi$_{1-x}$Mg$_x$O$_5$ (from Ref. 5) 
are reported. Through Eq. (\ref{rison}) in the text, the local magnetization of 
the classical chain can be quantitatively compared to the NMR spectra of the
$S$=1 chain and the correspondence between boundary moments and satellite
peaks is directly inferred.}
\label{fig2}
\end{figure*}


The purpose of the present work is to study whether or not the alternating
magnetization
observed in finite $S=1$ chains is a pure quantum mechanical effect somewhat
reminiscent
of the liquid-like ground state. The response of open chains with various
lengths $L$ to
a uniform magnetic field is thus studied in the limit of infinite-$S$,
proving that {\it
edge staggered defects with spin-$S/2$ originate from the translational
invariance
breaking, even in the framework of a classical model}. It is shown that the
characteristic spatial extention of these
defects corresponds to the zero-field correlation length of the
thermodynamic chain
($L\rightarrow\infty$) and that, at high temperature, the behavior of
$\langle S_{i}^{z}\rangle$
calculated for infinite-$S$ tracks the experimental findings in the $S=1$
chain.



In the classical limit ($S\rightarrow \infty$), the Hamiltonian for a
nearest-neighbor Heisenberg chain
of $L$ sites in an external field $H$ takes the form
\begin{equation}
{\cal H} = J \sum_{i=1}^{L} \vec S_i  \cdot \vec S_{i+1}-g\mu_B
H\sum_{i=1}^{L} S_i^z \; ,
\label{ham1.eq}
\end{equation}
where $ \vec S_i$'s are classical vectors, whose magnitude is taken to be
  $\left| \vec S_i \right| = \sqrt{S \left( S+1 \right) }$ (i.e
  $\left| \vec S_i \right| = \sqrt{2}$ in this case), and
$J$ is positive for antiferromagnets.



The thermal expectation value of the on-site magnetic moment has been
calculated by
the Wolff\cite{Wolff} cluster algorithm in the temperature range $50K\leq T
\leq 285K $,
for $L$ up to 31 spins. In order to compare numerical data with NMR results
in doped
Y$_{2}$BaNiO$_{5}$, $J/k_B$ has been set equal to 285 K \cite{Darriet}. The
effect of the
magnetic field $H$ has been included using a  standard Metropolis
algorithms, which
associates a flip probability to the whole cluster. As the field intensity
is low, the
flip acceptance remains close to one.


Typical results of the simulations for different values of $H$ are shown in
Fig. \ref{fig1}, proving that the response of a finite classical chain to a
homogeneous field is non uniform. In Fig. \ref{fig2}, magnetization profiles
at
characteristic temperatures and $H$=14.1 Tesla, are shown together with
$^{89}$Y
NMR spectra in Mg-doped Y$_{2}$BaNiO$_{5}$ (at the same temperatures) from Ref. 5.
The
spectra were obtained at fixed frequency ($ \nu_{rf}$ =29.4 MHz) by sweeping
the
magnetic field $H_0$ in a narrow range around 14.1 Tesla. The intensity of
the NMR
signal is proportional to the number of $^{89}$Y nuclei which obey the
resonance
condition
\begin{equation}
   H_0=2\pi \frac{\nu_{rf}-k}{\gamma}- A \langle S_i^{z} \rangle  \, \,
,
\label{rison}
\end{equation}
where $\gamma$ is the $^{89}$Y-gyromagnetic factor, $A=$1.3 Tesla is  the
$^{89}$Y - Ni$^{2+}$ hyperfine coupling constant in  Y$_{2}$BaNiO$_{5}$ and
$k$ is an $i$-independent factor that accounts for the chemical and orbital shifts. Using Eq.
(\ref{rison}), the local magnetization of the open classical chain and the
position of the peaks in the NMR spectra can be directly compared.
Since $k$ is not known precisely a priori, we fix it by matching the 
spin-polarization at the centre of the chain with the maximum of the NMR 
central line\cite{note1}. Then, as sketched in Fig. \ref{fig2}, a satellite peak is 
found in correspondence of each value of magnetization taken by the edge spins.



The magnetization profiles can be analyzed in detail, by observing that
$\langle S_i^z \rangle $ consists in a uniform part and of staggered
contributions decaying away from each boundary. The following
phenomenological function
\begin{equation}
\langle S_{i}^{z}\rangle=\langle S_{b}^{z}\rangle+\langle S_{1}^{z}\rangle
\, \left[ (-1)^{(i-1)}e^{\frac{(i-1)}{\xi}}+(-1)^{(L-i)}e^{\frac{(L-i)}{\xi}} \right]  \, 
\label{model.eq}
\end{equation}
is thus used to fit the numerical data. Expression (\ref{model.eq}) describes very 
well the behavior of $\langle S_i^z \rangle $ in all the investigated temperature 
range (see for instance solid lines in Fig. \ref{fig2}) and the extracted fitting 
parameters - $\langle S_{b}^{z}\rangle$, $\langle S_{1}^{z}\rangle$ and $\xi$ - 
turn out to be independent of the chain length $L$. The two alternating terms
in Eq. (\ref{model.eq}) give an in phase (out of phase) contribution in chains
with odd (even) number of spins. In real systems, consisting in an ensamble of
segments with different length, such effect induces a distribution of
$\langle S_i^z \rangle $ and thus a broadening of the NMR satellite peaks,
as actually observed in Fig. \ref{fig2}. Here we do not investigate this
subject quantitatively, being mainly interested to discuss the behavior of
the parameters extracted from the classical Monte Carlo data in light of the
values for the same quantities obtained by NMR in the $S$=1 chain.


\begin{figure}[b!]
\vspace{-10mm}
\begin{center}
\epsfxsize=87mm
 $$\epsffile{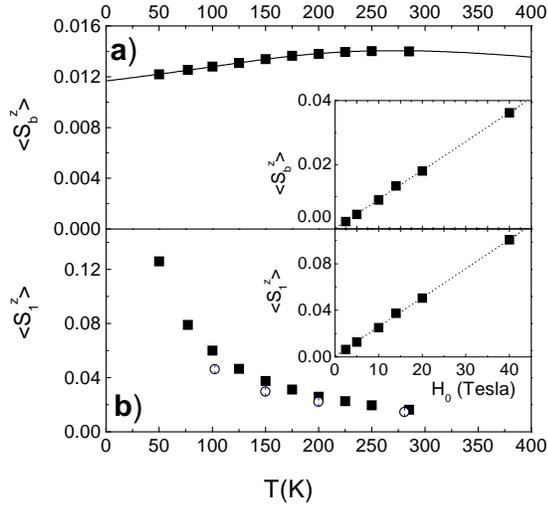}$$
\vspace{-5.8cm}
\caption{a) Temperature behavior of the uniform magnetization $\langle
S_b^z\rangle$ in $H=$14.1 Tesla extracted from classical Monte Carlo data 
by Eq. (\ref{model.eq}) (full squares). Solid line is the exact classical 
thermal expectation value of the on-site magnetic moment for a chain with 
$L \rightarrow \infty$, calculated through Fisher's formula for the uniform 
susceptibility \cite{Fisher}. b) The first-site staggered
magnetization
$\langle S_1^z \rangle$ of the classical chain is compared to the
experimental value
of $\langle S_1^z \rangle$ measured by NMR in Mg-doped Y$_{2}$BaNiO$_{5}$
\cite{Tedoldi}
(open circles). In the insets it is shown how $\langle S_b^z\rangle$ and
$\langle S_1^z\rangle$
depend linearly on the applied magnetic field up to 40 Tesla.}
\label{fig3}
\end{center}
\end{figure}




\begin{figure}[t!]
\vspace{-10mm}
\begin{center}
\epsfxsize=85mm
 $$\epsffile{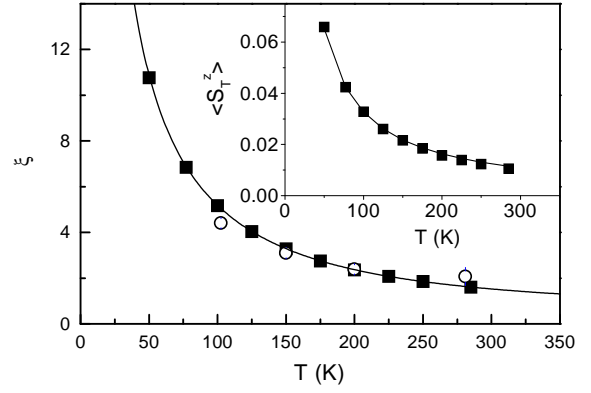}$$
\vspace{-7.0cm}
\caption{Decay length $\xi$ of the staggered magnetization, obtained
from Monte Carlo calculations through Eq. (\ref{model.eq}) (full squares),
{\it vis a vis} to the experimental values of $\xi$ (open
circles)\cite{Tedoldi}
and to the zero field correlation length of the infinite-$L$ classical
chain\cite{Fisher} (solid line). In inset, the total magnetization
$\langle S_T^z \rangle$, associated to one of the chain boundaries (see
Eq. (\ref{sum}) in the text), is fitted by the Curie-law (\ref{Curie}).}
\label{fig4}
\end{center}
\end{figure}



In Fig. \ref{fig3}a), the temperature behavior of the uniform term
$\langle S_{b}^{z}\rangle$ is reported. $\langle S_{b}^{z}\rangle$
depends linearly on the field up to 40 Tesla (inset) and follows
strictly the magnetization $\chi \cdot H $ that would be obtained by
using for $\chi$ the zero-field susceptibility of the infinite classical
chain\cite{Fisher} (solid line in Fig. \ref{fig3}a)). This result should
be intuitively expected, since the staggered terms in Eq. (\ref{model.eq})
involve only boundary spins and thus, in the thermodynamic limit, the
"bulk" magnetization is due only to $\langle S_{b}^{z}\rangle$. The first-site
magnitude $\langle S_{1}^{z}\rangle$ and the decay-length $\xi$ of the staggered
contributions, extracted from the simulation, are reported respectively
in Fig. \ref{fig3}b) and \ref{fig4} and compared with the experimental
values of $\langle S_{1}^{z}\rangle$ and $\xi$ from $^{89}$Y NMR in doped
Y$_{2}$BaNiO$_{5}$ \cite{Tedoldi}. The agreement between experimental
results in the HAF $S$=1 one-dimensional compound and predictions of
the classical model is remarkable at high temperature. Small quantitative
deviations, which likely prelude to a more sensible departure at lower 
temperatures, are observed only around $T=$100K$\simeq\Delta_H/k_B$. At the moment,
the lack of experimental data for $\langle S_{1}^{z}\rangle$ and $\xi$ below
100 K (due to the broadening of the NMR lines) prevents a precise definition
of the temperature range in which the classical model accounts for the
magnetic response of the open $S=1$ chain. Fig. \ref{fig4} also shows how the 
decay length $\xi$ of the staggered magnetization in a weak external field ($H$=14.1
Tesla) tracks strictly the behavior of the zero-field spin-spin correlation length (solid
line), calculated by Fisher\cite{Fisher} for the infinite-volume chain.


In the inset of Fig. \ref{fig4}, the total magnetic moment associated to
each staggered
contribution,
\begin{equation}
\langle S_T^z \rangle=\sum_{i=1}^L{\langle S_{1}^{z}\rangle
(-1)^{(i-1)}e^{\frac{(i-1)}{\xi}} \; ,
}
\label{sum}
\end{equation}
is plotted as a function of temperature. $\langle S_T^z \rangle$ is well
reproduced by a
Curie-like law
\begin{equation}
\langle S_T^z \rangle=\frac{g\mu_B S_f^2 H}{3k_BT} \; ,
\label{Curie}
\end{equation}
characteristic of non-interacting classical moments of magnitude $S_f$. The
value of $S_f$ that optimizes the fitting (solid line in the inset in Fig.
\ref{fig4}) is 0.72, about half of the site-spin magnitude $\sqrt{S \left( S+1 \right)}%
=\sqrt 2$. This result extends to a temperature region in which the system
shows classical behavior the low-$T$ picture for gapped spin chains, in which
$S$/2 degrees of freedom develop at the chain edges\cite{Hagiwara,Miya,White,Sore1,Polizzi}.
Moreover, since the temperature dependence of $\langle S_T^z \rangle$ is
mainly controlled by $\langle S_1^z \rangle$ for $\Delta_H /k_B\leq T \leq J/k_B$, even the
first-site staggered moment displays approximatively a Curie-like behavior.






In summary, our numerical results prove that the alternate boundary
magnetization observed experimentally in finite-length $S=1$ HAF chains
is correctly predicted, for $\Delta_H /k_B \leq T \leq J/k_B$,
by a strictly classical analysis of the Heisenbeg model. In
light of this evidence we conclude that the occurrence of Curie-like
fractional-spin defects is a general feature in HAF one-dimensional systems
with open boundaries, not related to the occurrence of a spin gap.
On the other hand, the size of these staggered defects, which is fixed by
the
bulk spin-spin correlation length, is affected by quantum fluctuations and
is
thus expected to remain finite when the ground state is spin-liquid like.






The authors are grateful to A. Rigamonti, E. S\o rensen and M. Fabrizio for
stimulating
discussions.










\begin{references}


\bibitem{Bobroff}    J. Bobroff {\it et al.}, Phys. Rev. Lett. {\bf 79}, 2117 (1997) and {\bf 80}, 3663 (1998).


\bibitem{Julien}     M. H. Julien {\it et al.}, \prl {\bf 84}, 3422 (2000).


\bibitem{Fujiwara}   N. Fujiwara {\it et al.}, Phys. Rev. Lett. {\bf 80}, 604 (1998).


\bibitem{Takigawa}   M. Takigawa, M.Motoyama. H.Eisaki and S. Uchida, Phys. Rev. B {\bf 55}, 14129 (1997).


\bibitem{Tedoldi}    F. Tedoldi, R. Santachiara and M. Horvati\'{c}, \prl {\bf 83}, 412 (1999).


\bibitem{Kennedy1}   T. Kennedy, J. Phys. Condens. Matter {\bf 2}, 5737 (1990).


\bibitem{Kennedy2}   T. Kennedy and H. Tasaki, Commun. Math. Phys. {\bf 147}, 431 (1992).


\bibitem{Miya}       S. Miyashita and S. Yamamoto, Phys. Rev. B {\bf 48}, 913 (1993).


\bibitem{White}      S. R. White and D. A. Huse, Phys. Rev. B {\bf 48}, 3844 (1993).


\bibitem{Sore1}      E. S. S\o rensen and I. Affleck, Phys. Rev. B {\bf 49},
15771 (1994).


\bibitem{Polizzi}    E. Polizzi, F. Mila, and E. S. S\o rensen, Phys. Rev. B {\bf 58}, 2407 (1998).


\bibitem{Alect}      F. Alet and E. S. S\o rensen, cond-mat/0006282.


\bibitem{Wolff}      U. Wolff, \prl {\bf 62}, 361 (1989).


\bibitem{Darriet}    J. Darriet and L. P. Regnault, Solid State Comm. {\bf 86}, 409 (1993).


\bibitem{note1} Assuming $\gamma$=$2\pi \cdot $2.0859 MHz/Tesla, the shift of the NMR central line
due to chemical and orbital contributions, extracted from the comparison between 
numerical and experimental results at $T=$ 283 K, turns out $k/\nu_{RF}=$390ppm, in good
agreement with previous extimations in the pure compound\cite{tedoriga,shimi}.


\bibitem{tedoriga}     F. Tedoldi and A. Rigamonti, Physica B {\bf259-261}, 995 (1999).


\bibitem{shimi}        T Shimizu {\it et al.}, Phys. Rev. B {\bf 52}, R9835 (1995).


\bibitem{Fisher}       M. E. Fisher, Am. J. Phys. {\bf 32}, 343 (1964).


\bibitem{Hagiwara}     M. Hagiwara {\it et al.}, \prl {\bf 65}, 3181 (1990).


\end{references}
\end{document}